\documentclass[12pt]{article}
\usepackage{epsfig}
\begin{document}
\begin{center}
{\Large {\bf The fluctuational region on the phase diagram of lattice Weinberg
- Salam model}

\vskip-40mm \rightline{\small ITEP-LAT/2009-10 } \vskip 30mm

{
\vspace{1cm}
{ M.A.~Zubkov$^{a,b}$ }\\
\vspace{.5cm} {\it $^a$ ITEP, B.Cheremushkinskaya 25, Moscow, 117259, Russia }
\\ \vspace{.5cm} {\it $^b$ Moscow Institute of Physics and Technology, 141700,
Dolgoprudnyi, Moscow Region, Russia  }}}

\end{center}

\begin{abstract}
The lattice Weinberg - Salam model without fermions is investigated numerically
for the realistic choice of bare coupling constants correspondent to the value
of the Weinberg angle $\theta_W \sim 30^o$, and the fine structure constant
$\alpha \sim \frac{1}{100}$. On the phase diagram there exists the vicinity of
the phase transition between the physical Higgs phase and the unphysical
symmetric phase, where the fluctuations of the scalar field become strong. The
classical Nambu monopole can be considered as an embryo of the unphysical
symmetric phase within the physical phase. In the fluctuational region quantum
Nambu monopoles are dense and, therefore, the perturbation expansion around
trivial vacuum cannot be applied. The maximal value of the cutoff at the given
values of coupling constants calculated using the lattices of sizes $8^3\times
16$, $12^3\times 16$, and  $16^4$ is $\Lambda_c \sim 1.4 \pm 0.2$ Tev. As the
lattice sizes used are rather small we consider this result as preliminary.

\end{abstract}

\section{Introduction}

In some phenomenological models that describe condensed matter
systems\footnote{One of the examples of such models is the Ginzburg - Landau
theory of superconductivity.} there exists the vicinity of the finite
temperature phase transition that is called fluctuational region. In this
region the fluctuations of the order parameter become strong. The contribution
of these fluctuations to certain physical observables becomes larger than the
tree level estimate. Thus the perturbation theory in these models fails down
within the fluctuational region.

Our main supposition is that the lattice Weinberg - Salam model (at $T = 0$)
looks similar to the mentioned models. Namely, we expect that there exists the
vicinity of the phase transition between the Higgs phase and the symmetric
phase in the Weinberg - Salam model, where the fluctuations of the scalar field
become strong and the perturbation expansion around trivial vacuum cannot be
applied. According to the numerical results the continuum theory is to be
approached within the vicinity of the phase transition, i.e. the cutoff is
increased along the line of constant physics when one approaches the point of
the transition. That's why we expect that the conventional prediction on the
value of the cutoff admitted in the Standard Model based on the perturbation
theory may be incorrect.

 According to the conventional point of view the upper bound $\Lambda$ on the cutoff in the
Electroweak theory (without fermions) depends on the Higgs mass. It is
decreased when the Higgs mass is increased. And at the Higgs mass around $1$
Tev $\Lambda$ becomes of the order of $M_H$. At the same time for $M_H \sim
200$ Gev the value of $\Lambda$ can be made almost infinite\footnote{Here we do
not consider vacuum stability bound on the Higgs mass related to the fermion
loops.}. This conclusion is made basing on the perturbation expansion around
trivial vacuum.

In the present paper we report the results of the numerical investigation of
the model at the value of the scalar self coupling $\lambda = 0.009$, the bare
Weinberg angle $\theta_W = 30^o$, and the renormalized fine structure constant
around $1/100$. The bare value of the Higgs boson mass is around $270$ Gev in
the vicinity of the phase transition.

We calculate the constraint effective potential $V(|\Phi|)$ for the Higgs field
$\Phi$. In the physical Higgs phase this potential has a minimum at a certain
nonzero value $\phi_m$ of $|\Phi|$. This shows that the spontaneous breakdown
of the Electroweak symmetry takes place as it should. However, there exists the
vicinity of the phase transition, where the fluctuations of the Higgs field are
of the order of $\phi_m$ while the hight of the "potential
barrier"\footnote{The meaning of the words "potential barrier" here is
different from that of the one - dimensional quantum mechanics as here
different minima of the potential form the three - dimensional sphere while in
usual $1D$ quantum mechanics with the similar potential there are two separated
minima with the potential barrier between them. Nevertheless we feel it
appropriate to use the chosen terminology as the value of the "potential
barrier hight" measures the difference between the potentials with and without
spontaneous symmetry breaking. }  $H = V(0) - V(\phi_m)$ is of the order of
$V(\phi_m + \delta \phi)-V(\phi_m)$, where $\delta \phi$ is the fluctuation of
$|\Phi|$. We expect that in this region the perturbation expansion around
trivial vacuum $\Phi = (\phi_m,0)^T$ cannot be applied. We call this region of
the phase diagram the fluctuational region (FR) in analogy to the condensed
matter systems.

The mentioned supposition is confirmed by the investigation of the topological
defects composed of the lattice gauge fields that are to be identified with
quantum Nambu monopoles \cite{Nambu,BVZ,Chernodub_Nambu}. We show that their
lattice density increases when the phase transition point is approached. Within
the FR these objects are so dense that it is not possible at all to speak of
them as of single monopoles \footnote{It has been shown in \cite{VZ2008} that
at the infinite value of the scalar self coupling $\lambda = \infty$ moving
along the line of constant physics we reach the point on the phase diagram
where the monopole worldlines begin to percolate. This point was found to
coincide roughly with the position of the transition between the physical Higgs
phase and the unphysical symmetric phase of the lattice model. This transition
is a crossover and the ultraviolet cutoff achieves its maximal value around
$1.4$ Tev at the transition  point.}. Namely, within this region the average
distance between the Nambu monopoles is of the order of their size. Such
complicated configurations obviously have nothing to do with the conventional
vacuum used of the continuum perturbation theory.

We have estimated the maximal value of the cutoff in the vicinity of the
transition point. The obtained value of the cutoff appears to be around $1.4$
Tev.

\section{The lattice model under investigation}
The lattice Weinberg - Salam Model without fermions contains  gauge field
${\cal U} = (U, \theta)$ (where $ \quad U
 \in SU(2), \quad e^{i\theta} \in U(1)$ are
realized as link variables), and the scalar doublet $ \Phi_{\alpha}, \;(\alpha
= 1,2)$ defined on sites.

The  action is taken in the form
\begin{eqnarray}
 S & = & \beta \!\! \sum_{\rm plaquettes}\!\!
 ((1-\mbox{${\small \frac{1}{2}}$} \, {\rm Tr}\, U_p )
 + \frac{1}{{\rm tg}^2 \theta_W} (1-\cos \theta_p))+\nonumber\\
 && - \gamma \sum_{xy} Re(\Phi^+U_{xy} e^{i\theta_{xy}}\Phi) + \sum_x (|\Phi_x|^2 +
 \lambda(|\Phi_x|^2-1)^2), \label{S}
\end{eqnarray}
where the plaquette variables are defined as $U_p = U_{xy} U_{yz} U_{wz}^*
U_{xw}^*$, and $\theta_p = \theta_{xy} + \theta_{yz} - \theta_{wz} -
\theta_{xw}$ for the plaquette composed of the vertices $x,y,z,w$. Here
$\lambda$ is the scalar self coupling, and $\gamma = 2\kappa$, where $\kappa$
corresponds to the constant used in the investigations of the $SU(2)$ gauge
Higgs model. $\theta_W$ is the Weinberg angle. Bare fine structure
 constant $\alpha$ is expressed through $\beta$ and $\theta_W$ as $\alpha = \frac{{\rm tg}^2 \theta_W}{\pi \beta(1+{\rm tg}^2
\theta_W)}$. In our investigation we fix bare  Weinberg angle equal to $30^o$.
The renormalized fine structure constant can be extracted through the potential
for the infinitely heavy external charged particles.

All simulations were performed on lattices of sizes $8^3\times 16$ and
$12^3\times 16$. Several points were checked using the larger lattice ($16^4$).

In order to simulate the system we used Metropolis algorithm. The acceptance
rate is kept around $0.5$ via the automatical self - tuning of the suggested
distribution of the fields. At each step of the suggestion the random value is
added to the old value of the scalar field while the old value of Gauge field
is multiplied by random $SU(2)\otimes U(1)$ matrix. We use Gaussian
distribution both for the random value added to the scalar field and the
parameters of the random matrix multiplied by the lattice Gauge field. We use
two independent parameters for these distributions: one for the Gauge fields
and another for the scalar field. The program code has been tested for the case
of frozen scalar field. And the results of the papers \cite{BVZ} are repeated.
We also have tested our code for the $U(1)$ field frozen and repeat the results
of \cite{Montvayold}. For the values of couplings used on the lattice $16^4$
the autocorrelation time for the gauge fields is estimated as about $N^g_{auto}
\sim 2500$ Metropolis steps. (The correlation between the values of the gauge
field is less than $3 \%$ for the configurations separated by $N^g_{auto}$
Metropolis steps. Each metropolis step consists of the renewing the fields over
all the lattice.) The autocorrelation time for the scalar field is much less
$N^{\phi}_{auto} \sim 20$. The estimated time for the preparing the equilibrium
starting from the cold start is about $18000$ Metropolis steps.

\begin{figure}
\begin{center}
 \epsfig{figure=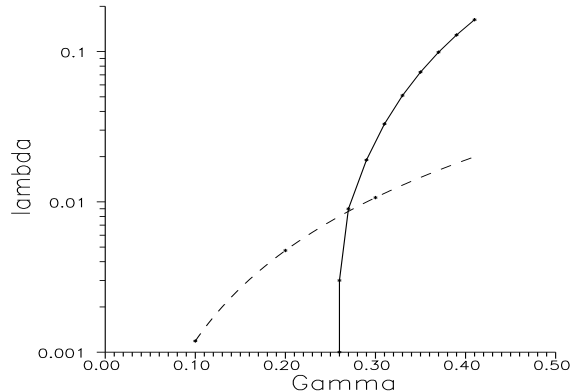,height=60mm,width=80mm,angle=0}
\caption{\label{fig.2} The phase diagram of the model in the
 $(\gamma, \lambda)$-plane at $\beta = 12$. The dashed line is the tree - level
  estimate for the line of constant physics correspondent to bare $M^0_H = 270$ Gev. The continuous line
  is the line of phase transition between the physical Higgs phase and the unphysical symmetric phase.   }
\end{center}
\end{figure}

\section{Phase diagram}

 In the three - dimensional ($\beta, \gamma, \lambda$) phase
diagram the transition surfaces are two - dimensional. The lines of constant
physics on the tree level are the lines ($\frac{\lambda}{\gamma^2} = \frac{1}{8
\beta} \frac{M^2_H}{M^2_W} = {\rm const}$; $\beta = \frac{1}{4\pi \alpha}={\rm
const}$). We suppose that in the vicinity of the transition  the deviation of
the lines of constant physics from the tree level estimate may be significant.
However,  qualitatively their behavior is the same. Namely, the cutoff is
increased along the line of constant physics when $\gamma$ is decreased and the
maximal value of the cutoff is achieved at the transition point. Nambu monopole
density in lattice units is also increased when the ultraviolet cutoff is
increased.

At $\beta = 12$ the phase diagram is represented on Figure \ref{fig.2}. The
physical Higgs phase is situated right to the transition line. The position of
the transition is localized at the point where the susceptibility extracted
from the Higgs field creation operator achieves its maximum. The following
variable is considered as creating the $Z$ boson:
\begin{equation} Z_{xy} = Z^{\mu}_{x} \;
 = {\rm sin} \,[{\rm Arg} (\Phi_x^+U_{xy} e^{i\theta_{xy}}\Phi_y) ]. \label{Z1}
\end{equation}
 We use the
susceptibility  $\chi = \langle H^2 \rangle - \langle H\rangle^2$ extracted
from $H = \sum_{y} Z^2_{xy}$. We observe no difference between the values of
the susceptibility calculated using the lattices of the sizes $8^3\times 16$,
$12^3\times 16$, and $16^4$. This indicates that the transition may be a
crossover.

It is worth mentioning that the value of the renormalized Higgs boson mass does
not deviate significantly from its bare value. Namely, for $\lambda$ around
$0.009$ and $\gamma$ in the vicinity of the phase transition  bare value of the
Higgs mass is around $270$ Gev while the observed renormalized value is
 $300 \pm 70$ Gev (see the next section for the details).

\section{Masses and the lattice spacing}

In order to evaluate the masses of the $Z$-boson and the Higgs boson we use the
correlators:
\begin{equation}
\frac{1}{N^6} \sum_{\bar{x},\bar{y}} \langle \sum_{\mu} Z^{\mu}_{x} Z^{\mu}_{y}
\rangle   \sim
  e^{-M_{Z}|x_0-y_0|}+ e^{-M_{Z}(L - |x_0-y_0|)}
\label{corZ}
\end{equation}
and
\begin{equation}
  \frac{1}{N^6}\sum_{\bar{x},\bar{y}}(\langle H_{x} H_{y}\rangle - \langle H\rangle^2)
   \sim
  e^{-M_{H}|x_0-y_0|}+ e^{-M_{H}(L - |x_0-y_0|)},
\label{cor}
\end{equation}
 Here the summation $\sum_{\bar{x},\bar{y}}$ is over the three ``space"
components of the four - vectors $x$ and $y$ while $x_0, y_0$ denote their
``time" components. $N$ is the lattice length in "space" direction. $L$ is the
lattice length in the "time" direction.

In lattice calculations we used two different operators that create Higgs
bosons: $ H_x = |\Phi|$ and $H_x = \sum_{y} Z^2_{xy}$. In both cases $H_x$ is
defined at the site $x$, the sum $\sum_y$ is over its neighboring sites $y$.

After fixing the unitary gauge ($\Phi_2 = 0$; $\Phi_1 \in {\cal R}$; $\Phi_1
\ge 0$), lattice Electroweak theory becomes a lattice $U(1)$ gauge theory with
the $U(1)$ gauge field
\begin{equation}
 A_{xy}  =  A^{\mu}_{x} \;
 = \,[-Z^{\prime}  + 2\theta_{xy}]  \,{\rm mod}
 \,2\pi, \label{A}
\end{equation}
where the new lattice $Z$ - boson field (different from (\ref{Z1})) is defined
as
\begin{equation}
Z^{\prime} =  {\rm Arg} (\Phi_x^+U_{xy} e^{i\theta_{xy}}\Phi_y) .\label{Z2}
\end{equation}

The usual Electromagnetic field is related to $A$ as $ A_{\rm EM}  =  A +
Z^{\prime} - 2 \,{\rm sin}^2\, \theta_W Z^{\prime}$.

The physical scale is given in our lattice theory by the value of the $Z$-boson
mass $M^{phys}_Z \sim 91$ GeV. Therefore the lattice spacing is evaluated to be
$a \sim [91 {\rm GeV}]^{-1} M_Z$, where $M_Z$ is the $Z$ boson mass in lattice
units. The similar calculations have been performed in \cite{VZ2008} for
$\lambda = \infty$. It has been shown that the $W$ - boson mass unlike $M_Z$
depends strongly on the lattice size due the photon cloud. Therefore the $Z$ -
boson mass was used in \cite{VZ2008} in order to fix the scale. That's why in
the present paper we do not consider the $W$ - boson mass.

Our data obtained on the lattice $8^3\times16$ shows that $\Lambda=
\frac{\pi}{a} = (\pi \times 91~{\rm GeV})/M_Z$ is increased slowly with the
decrease of $\gamma$ at any fixed $\lambda$. We investigated carefully the
vicinity of the transition point at fixed $\lambda = 0.009$ and $\beta = 12$.
It has been found that at the transition point $\gamma_c = 0.273 \pm 0.002$ the
value of $\Lambda$ is equal to $1.4 \pm 0.2$ Tev. The check of a larger lattice
(of size $12^3\times 16$) does not show an essential increase of this value. We
also calculate $\Lambda$ on the lattice $16^4$ at the two points (one is at the
transition point and another is within the physical phase). Again we do not
observe the increase of $\Lambda$.
 However, at the present moment we do not exclude that such an increase can be observed on the larger lattices. That's
 why careful
investigation of the dependence of $\Lambda$ on the lattice size (as well as on
$\lambda$) must be performed in order to draw the final conclusion.
 On Fig. 3 the dependence of $M_Z$ in
lattice units on $\gamma$ is represented at $\lambda =0.009$ and $\beta = 12$.

In the Higgs channel the situation is much more difficult. First, due to the
lack of statistics we cannot estimate the masses in this channel using the
correlators (\ref{cor}) at all considered values of $\gamma$. At the present
moment we can represent the data at the two points on the lattice
$8^3\times16$: ($\gamma = 0.274$, $\lambda =0.009$, $\beta = 12$) and ($\gamma
= 0.290$, $\lambda =0.009$, $\beta = 12$). The first point roughly corresponds
to the position of the transition while the second point is situated deep
within the Higgs phase. The sets of coupling chosen correspond to bare Higgs
mass around $270$ Gev. That's why in this channel, in principle,  bound states
of the gauge bosons may appear. This situation was already considered in
earlier studies of $SU(2)$ Gauge - Higgs model (see, for example,
\cite{Montvay} and references therein). Following these studies we interpret
the mass found in this channel at small "time" separations as the Higgs mass.
We suppose that the bound states of gauge bosons may appear in correlator
(\ref{cor}) at larger "time" separations.

At the point ($\gamma = 0.274$, $\lambda =0.009$, $\beta = 12$) we have
collected enough statistics to calculate correlator (\ref{cor}) up to the
"time" separation $|x_0-y_0| = 4$. The value $\gamma = 0.274$ corresponds
roughly to the position of the phase transition. The mass found in this channel
in lattice units is $M^L_H = 0.75 \pm 0.1$ while bare value of $M_H$ is $M^0_H
\sim 270$ Gev. At the same time $M_Z^L = 0.23 \pm 0.007$. Thus we estimate at
this point $M_H = 300 \pm 40$ Gev.

At the point ($\gamma = 0.29$, $\lambda =0.009$, $\beta = 12$) we calculate the
correlator with reasonable accuracy up to $|x_0-y_0| = 3$.  At this point bare
value of $M_H$ is $M^0_H \sim 260$ Gev while the renormalized Higgs mass in
lattice units is $M^L_H = 1.2 \pm 0.3$. At the same time $M_Z^L = 0.41 \pm
0.01$. Thus we estimate at this point $M_H = 265 \pm 70$ Gev.

It is worth mentioning that in order to calculate $Z$ - boson mass we fit the
correlator (\ref{corZ}) for $8 \ge |x_0-y_0| \ge 1$. In order to calculate the
Higgs boson mass at $\gamma = 0.274$ we use the data for the correlator
(\ref{cor}) at $4 \ge |x_0-y_0| \ge 0$. In order to calculate the Higgs boson
mass at $\gamma = 0.29$ we use the correlator for $3 \ge |x_0-y_0| \ge 0$.

\begin{figure}
\begin{center}
 \epsfig{figure=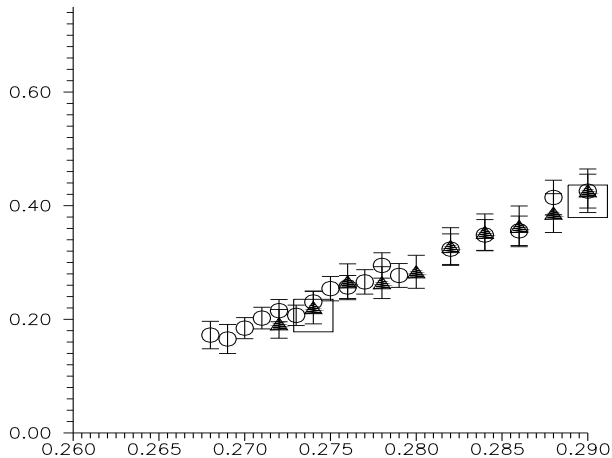,height=60mm,width=80mm,angle=0}
\caption{\label{fig.3} Z - boson mass in lattice units at $\lambda =0.009$ and
$\beta = 12$. Circles correspond to lattice $8^3\times 16$. Triangles
correspond to lattice $12^3\times 16$. Squares correspond to lattice $16^4$
(the error bars are about of the same size as the symbols used).}
\end{center}
\end{figure}

\section{Effective constraint potential}

\begin{figure}
\begin{center}
 \epsfig{figure=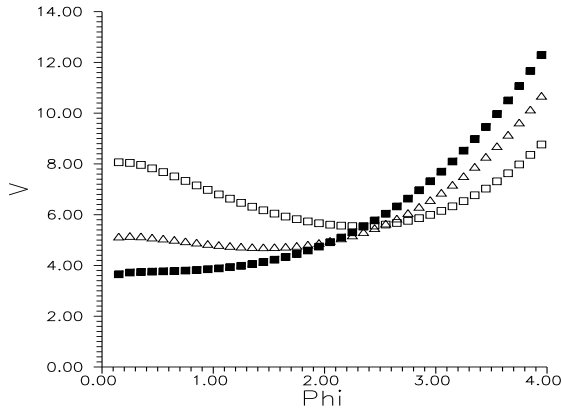,height=60mm,width=80mm,angle=0}
\caption{\label{fig.1} The effective constraint potential at $\lambda =0.009$
and $\beta = 12$. Black squares correspond to $\gamma_c = 0.273$. Empty squares
correspond to $\gamma =0.29$. Triangles correspond to $\gamma = 0.279$. The
error bars are about of the same size as the symbols used. }
\end{center}
\end{figure}

We have calculated the constraint effective potential for $|\Phi|$ using the
histogram method. The calculations have been performed on the lattice
$8^3\times 16$. The probability $h(\phi)$ to find the value of $|\Phi|$ within
the interval $[\phi-0.05;\phi+0.05)$ has been calculated for $\phi = 0.05 +
N*0.1$, $N = 0,1,2, ...$ This probability is related to the effective potential
as $ h(\phi) = \phi^3 e^{-V(\phi)}$. That's why we extract the potential from
$h(\phi)$ as
\begin{equation}
V(\phi) = - {\rm log}\, h(\phi) + 3 \, {\rm log} \, \phi \label{CEP}
\end{equation}
It is worth mentioning that $h(0.05)$ is calculated as the probability to find
the value of $|\Phi|$ within the interval $[0;0.1]$. Within this interval ${\rm
log}\, \phi$ is ill defined. That's why we exclude the point $\phi = 0.05$ from
our data. Instead we calculate $V(0)$ using the extrapolation of the data at
$0.15 \le \phi \le 2.0$. The extrapolation is performed using the polynomial
fit with the powers of $\phi$ up to the third (average deviation of the fit
from the data is around $1$ per cent). Next, we introduce the useful quantity
$H = V(0) - V(\phi_m)$, which is called the potential barrier hight (here
$\phi_m$ is the point, where $V$ achieves its minimum).

\begin{table}
\label{Table}\caption{The values of $\phi_m$, $H$, $H_{\rm fluct}$, and Nambu
monopole density $\rho$  at selected values of $\gamma$ for $\lambda = 0.009$,
$\beta = 12$ (Lattice $8^3\times 16$.)}
\begin{center}
\begin{tabular}{|c|c|c|c|c|c|}
\hline
{\bf $\gamma$}  & {\bf $\phi_m$}  & {\bf $H$}& $H_{\rm fluct}$& $\rho$ \\
\hline $0.273$  & $0$ & $0$ & $0.1\pm 0.1$ & $0.098\pm 0.001$\\
\hline $0.274$  & $0$ & $0$ & $0.04 \pm 0.1$ & $0.081\pm 0.001$\\
\hline $0.275$  & $0.85\pm 0.1$ & $0.01 \pm 0.06$&  $0.15\pm 0.05$ & $0.067\pm 0.001$\\
\hline $0.276$  & $1.05\pm 0.1$ & $0.05\pm 0.06$& $0.16\pm 0.01$ & $0.054\pm 0.001$\\
\hline $0.277$  & $ 1.25 \pm 0.05$ &$ 0.19\pm 0.05$ &$0.25\pm 0.05$ & $0.044\pm 0.001$ \\
\hline $0.278$  & $ 1.35 \pm 0.1$  & $0.28 \pm 0.07$&$0.25\pm 0.06$& $0.035\pm 0.001$\\
\hline $0.279$  & $1.45\pm 0.05$ & $0.5 \pm 0.06$&  $0.25\pm 0.06$& $0.028\pm 0.001$\\
\hline $0.282 $ & $1.75\pm 0.05$ & $1.04\pm 0.07$&$0.31\pm 0.07$& $0.014\pm 0.001$\\
\hline $0.284 $ & $1.95\pm 0.05$ & $1.41\pm 0.08$&$0.38\pm 0.08$& $0.0082\pm 0.0005$\\
\hline $0.286 $ & $2.05\pm 0.05$ & $1.86\pm 0.08$&$0.35\pm 0.08$& $0.0049\pm 0.0002$\\
\hline $0.288 $ & $2.15\pm 0.05$ & $2.33\pm 0.08$&$0.32\pm 0.07$& $0.0029\pm 0.0002$\\
\hline $0.29 $ & $2.25\pm 0.05$ & $2.82\pm 0.08$&$0.44\pm 0.08$ & $0.0017\pm 0.0001$\\
\hline
\end{tabular}
\end{center}
\end{table}

In Table $1$ we represent the values of $\phi_m$ and $H$ for $\lambda = 0.009$,
$\beta = 12$.  One can see that the values of $\phi_m$  and $H$ increase when
$\gamma$ is increased. At $\gamma = 0.273$ the minimum of the potential is at
$\phi = 0$. This point corresponds to the maximum of the susceptibility
constructed of the Higgs field creation operator. At $\gamma = 0.274$ we also
observe the only minimum for the potential at $\phi = 0$. At $\gamma = 0.275$
minimum of the potential is observed at $\phi_m = 0.85\pm 0.1$ with the very
small barrier hight. That's why we localize the position of the transition
point at $\gamma = 0.273\pm 0.002$.

It is important to understand which value of barrier hight can be considered as
small and which value can be considered as large. Our suggestion is to compare
$H = V(0) - V(\phi_m)$ with $H_{\rm fluct} = V(\phi_m + \delta \phi) -
V(\phi_m)$, where $\delta \phi$ is the fluctuation of $|\Phi|$.

From Table $1$ it is clear that there exists the value of $\gamma$ (we denote
it $\gamma_{c2}$) such that at $\gamma_c < \gamma < \gamma_{c2}$ the barrier
hight $H$ is of the order of $H_{\rm fluct}$ while for $\gamma_{c2} << \gamma$
the barrier hight is essentially larger than $H_{\rm fluct}$. The rough
estimate for this pseudocritical value is $\gamma_{c2} \sim 0.278$.

We estimate the fluctuations of $|\Phi|$ to be around $\delta \phi \sim 0.6$
for all considered values of $\gamma$ at $\lambda = 0.009$, $\beta = 12$. It
follows from our data that $\phi_m >> \delta \phi$ at $\gamma_{c2} << \gamma$
while $\phi_m \sim \delta \phi$ at $\gamma_{c2} > \gamma$.

Basing on these observations we expect that in the region $\gamma_{c2} <<
\gamma$ the usual perturbation expansion around trivial vacuum of spontaneously
broken theory can be applied to the lattice Weinberg - Salam model while in the
FR $\gamma_c < \gamma < \gamma_{c2}$ it cannot be applied.

At the value of $\gamma$ equal to $\gamma_{c2} \sim 0.278$ the calculated value
of the cutoff is $1.0 \pm 0.1 $ Tev.

\section{The renormalized coupling}

In order to calculate the renormalized fine structure constant $\alpha_R =
e^2/4\pi$ (where $e$ is the electric charge) we use the potential for
infinitely heavy external fermions. We consider Wilson loops for the
right-handed external leptons: $
 {\cal W}^{\rm R}_{\rm lept}(l)  =
 \langle {\rm Re} \,\Pi_{(xy) \in l} e^{2i\theta_{xy}}\rangle.
$
Here $l$ denotes a closed contour on the lattice. We consider the following
quantity constructed from the rectangular Wilson loop of size $r\times t$:
$
 {\cal V}(r) = \lim_{t \rightarrow \infty}{
 \rm log}
 \frac{  {\cal W}(r\times t)}{{\cal W}(r\times (t+1))}.
$
At large enough distances we expect the appearance of the Coulomb interaction
$
 {\cal V}(r) = -\frac{\alpha_R}{r} + const.
$

 At $\lambda = 0.009$, $\beta = 12$, $ \gamma = \gamma_c(\lambda)\sim 0.273$ the renormalized fine
structure constant calculated on the lattice $8^3\times 16$ is $\alpha_R =
\frac{1}{97 \pm 2}$. For $0.27 \le \gamma \le 0.29$ the value of $\alpha_R$
varies between $ \frac{1}{97 \pm 2}$ and $\frac{1}{100 \pm 2}$. These values
coincide within the statistical errors with the values calculated on the larger
lattice ($12^3\times 16$). This indicates that the value of $\alpha_R$ does not
depend on the lattice size also for the small values of $\lambda$. The
calculated values are to be compared with bare constant $\alpha_0 = 1/(4\pi
\beta)$.

\section{Nambu monopoles}

In this section we remind the reader what is called Nambu monopole \cite{Nambu,
BVZ}. First let us define the continuum Electroweak fields as they appear in
the Weinberg-Salam model. The continuum scalar doublet is denoted as $\Phi$.
After fixing the unitary gauge $\Phi_1 \in {\cal R}$, $\Phi_2 = 0$, the
$Z$-boson field $Z^{\mu}$ and electromagnetic field $A_{\rm EM}^{\mu}$ are
defined as $ Z^{\mu} = \frac{1}{2}{\rm Tr}C^{\mu}\sigma_3 +  B^{\mu}$, and
 $A_{\rm EM}^{\mu} =  2 B^{\mu}  - 2 \,{\rm sin}^2\, \theta_W Z^{\mu}$,
where $C^{\mu}$ and $B^{\mu}$ are the corresponding $SU(2)$ and $U(1)$ gauge
fields of the Standard Model.

Nambu monopoles are defined as the endpoints of the $Z$-string \cite{Nambu}.
The $Z$-string is the classical field configuration that represents the object,
which is characterized by the magnetic flux extracted from the $Z$-boson field.
Namely, for a small contour $\cal C$ winding around the $Z$ - string one should
have
\begin{equation}
 \int_{\cal C} Z^{\mu} dx^{\mu} \sim 2\pi;\,
 \int_{\cal C} A_{\rm EM}^{\mu} dx^{\mu} \sim 0;\,
 \int_{\cal C} B^{\mu} dx^{\mu} \sim 2\pi {\rm sin}^2\, \theta_W .
\end{equation}
The string terminates at the position of the Nambu monopole. The hypercharge
flux is supposed to be conserved at that point. Therefore, a Nambu monopole
carries electromagnetic flux $4\pi {\rm sin}^2\, \theta_W$. The size of Nambu
monopoles was estimated \cite{Nambu} to be of the order of the inverse Higgs
mass, while its mass should be of the order of a few TeV. According to
\cite{Nambu} Nambu monopoles may appear only in the form of a bound state of a
monopole-antimonopole pair.

In lattice theory the classical solution corresponding to a $Z$-string should
be formed around the $2$-dimensional topological defect which is represented by
the integer-valued field defined on the dual lattice $ \Sigma =
\frac{1}{2\pi}^*([d Z^{\prime}]_{{\rm mod} 2\pi} - d Z^{\prime})$. (Here we
used the notations of differential forms on the lattice. For a definition of
those notations see, for example, ~\cite{forms}. Lattice field $Z^\prime$ is
defined in Eg. (\ref{Z2}).) Therefore, $\Sigma$ can be treated as the
worldsheet of a {\it quantum} $Z$-string\cite{Chernodub_Nambu,BVZ}. Then, the
worldlines of quantum Nambu monopoles appear as the boundary of the $Z$-string
worldsheet: $ j_Z = \delta \Sigma $.

It has been mentioned in  section $5$ that our lattice model becomes $U(1)$
gauge model after fixing the unitary gauge. The corresponding compact $U(1)$
gauge field is given by Eq.~(\ref{A}). Therefore one may try to extract
monopole trajectories directly from $A$. The monopole current is given by
\begin{equation}
 j_{A} = \frac{1}{2\pi} {}^*d([d A]{\rm mod}2\pi)
\label{Am}
\end{equation}
Eg. (\ref{A}) in continuum notations is
\begin{equation}
 A^{\mu}  =  - Z^{\mu} + 2 B^{\mu},
\end{equation}
where $B$ is the hypercharge field. Its strength is divergenceless. As a result
in continuum theory the net $Z$ flux emanating from the center
 of the monopole is equal to the net $A$ flux with the opposite sign.
(Both $A$ and $Z$ are undefined inside the monopole.)  This means that in the
continuum limit the position of the Nambu monopole must coincide
 with the position of the antimonopole extracted from the field $A$.
Therefore, one can consider Eq.~(\ref{Am}) as another definition of a quantum
Nambu monopole \cite{BVZ,VZ2008}. Actually, in our numerical simulations we use
the definition of Eq. (\ref{Am}).

\section{Nambu monopole density }

According to the previous section  the worldlines of the quantum Nambu
monopoles can be extracted from the field configurations according to Eq.
(\ref{Am}).
The monopole density is defined as $ \rho = \left\langle \frac{\sum_{\rm
links}|j_{\rm link}|}{4V^L}
 \right\rangle,$
where $V^L$ is the lattice volume.

In Table $1$ we represent Nambu monopole density as a function of $\gamma$ at
$\lambda = 0.009$, $\beta = 12$.  The value of monopole density at $\gamma_c =
0.273$ is around $0.1$. At this point the value of the cutoff is $\Lambda \sim
1.4 \pm 0.2$ Tev.

According to the classical picture the Nambu monopole size is of the order of
$M^{-1}_H$. Therefore for $a^{-1} \sim 430$ Gev and $M_H \sim 300$ Gev the
expected size of the monopole is about $1.4$ lattice units.

The monopole density around $0.1$ means that among $10$ sites there exists $4$
sites that are occupied by the monopole. Average distance between the two
monopoles is, therefore, less than $1$ lattice spacing and it is not possible
at all to speak of the given configurations as of representing the physical
Nambu monopole.

At $\gamma = \gamma_{c2}\sim 0.278$ the Nambu monopole density is around
$0.035$. This means that among $7$ sites there exists one site that is occupied
by the monopole. Average distance between the two monopoles is, therefore,
approximately $2$ lattice spacings or $\sim \frac{1}{160\, {\rm Gev}}$. Thus,
the Nambu monopole density in physical units is around $[{160\, {\rm Gev}}]^3$.
We see that at this value of $\gamma$ the average distance between Nambu
monopoles is of the order of their size.

We summarize the above observations as follows. Within the fluctuational region
the configurations under consideration do not represent single Nambu monopoles.
Instead these configurations can be considered as the collection of monopole -
like objects that is so dense that the average distance between the objects is
of the order of their size. On the other hand, at $\gamma
>> \gamma_{c2}$ the considered configurations do represent single Nambu
monopoles and the average distance between them is much larger than their size.
In other words out of the FR vacuum can be treated as a gas of Nambu monopoles
while within the FR vacuum can be treated as a liquid composed of monopole -
like objects.

It is worth mentioning that somewhere inside the $Z$ string connecting the
classical Nambu monopoles the Higgs field is zero: $|\Phi| = 0$. This means
that the $Z$ string with the Nambu monopoles at its ends can be considered as
an embryo of the symmetric phase within the Higgs phase. We observe that the
density of these embryos is increased when the phase transition is approached.
Within the fluctuational region the two phases are mixed, which is related to
the large value of Nambu monopole density.

That's why we come to the conclusion that vacuum of lattice Weinberg - Salam
model within the FR has nothing to do with the continuum perturbation theory.
This means that the usual perturbation expansion around trivial vacuum (gauge
field equal to zero, the scalar field equal to $(\phi_m,0)^T$) cannot be valid
within the FR.  This might explain why we do not observe in our numerical
simulations the large values of $\Lambda$ predicted by the conventional
perturbation theory.

\section{Conclusions}

In the present paper we demonstrate that there exists the so - called
fluctuational region (FR) on the phase diagram of the lattice Weinberg - Salam
model. This region is situated in the vicinity of the phase transition between
the physical Higgs phase and the unphysical symmetric phase of the model. We
calculate the effective constraint potential $V(\phi)$ for the Higgs field. It
has a minimum at the nonzero value $\phi_m$ in the physical Higgs phase. Within
the FR the fluctuations of the scalar field become of the order of $\phi_m$.
Moreover, the barrier hight $H = V(0) - V(\phi_m)$ is of the order of $V(\phi_m
+ \delta \phi)- V(\phi_m)$, where $\delta \phi$ is the fluctuation of $|\Phi|$.

The scalar field must be equal to zero somewhere within the classical Nambu
monopole. That's why this object can be considered as an embryo of the
unphysical symmetric phase within the physical Higgs phase of the model. We
investigate properties of the quantum Nambu monopoles. Within the FR they are
so dense that the average distance between them becomes of the order of their
size. This means that the two phases are mixed within the FR.

All these results show that the vacuum of lattice Weinberg - Salam model in the
FR is essentially different from the trivial vacuum used in the conventional
perturbation theory. As a result the usual perturbation theory cannot be
applied in this region.

It is important that the continuum physics is to be approached in the vicinity
of the mentioned phase transition. The ultraviolet cutoff is increased when the
transition point is approached along the line of constant physics. Our
numerical results show that at $M_H$ around $300$ Gev and the fine structure
constant around $1/100$ the maximal value of the cutoff admitted out of the FR
for the considered lattice sizes cannot exceed the value around $1.0 \pm 0.1$
Tev. Within the FR the larger values of the cutoff can be achieved in principle
although we expect the correspondent results cannot be related directly to the
real continuum physics. The absolute maximum for the value of the cutoff within
the Higgs phase of the lattice model is achieved at the point of the phase
transition. Our estimate for this value is $1.4 \pm 0.2$ Tev for the lattice
sizes $8^3\times 16$, $12^3\times 16$, and $16^4$. Out of the fluctuational
region the behavior of the lattice model in general is close to what we expect
basing on the continuous perturbation theory.

We do not observe the dependence of the observables on the lattice size
($8^3\times 16$, $12^3\times 16$, and $16^4$). In particular, the
susceptibility extracted from the Higgs field creation operator does not depend
on the lattice size. This indicates that the transition at the considered value
of $\lambda = 0.009$ is a crossover. However, our results have been obtained on
rather small lattices. That's why careful investigation of the dependence of
the observables on the lattice size (as well as on the Higgs mass) must be
performed in order to draw the final conclusions. The inclusion of the
dynamical fermions into the consideration may also be important.

This work was partly supported by RFBR grants 09-02-00338, 08-02-00661, and
07-02-00237, by Grant for leading scientific schools 679.2008.2,by Federal
Program of the Russian Ministry of Industry, Science and Technology No
40.052.1.1.1112. The numerical simulations have been performed using the
facilities of Moscow Joint Supercomputer Center.


\begin{thebibliography}{99}

\bibitem{Nambu}
Y.~Nambu, Nucl.Phys. B {\bf 130}, 505 (1977);\\
Ana~Achucarro and Tanmay~Vachaspati, Phys. Rept. {\bf 327}, 347 (2000); Phys.
Rept. {\bf 327}, 427 (2000).

\bibitem{BVZ}
B.L.G.~Bakker, A.I.~Veselov, and M.A.~Zubkov, Phys. Lett. B {\bf  583}, 379
(2004); Yad. Fiz. {\bf 68}, 1045 (2005); Phys. Lett. B {\bf 620}, 156 (2005);
 Phys. Lett. B {\bf 642}, 147 (2006);  J. Phys. G: Nucl. Part. Phys. 36
(2009) 075008; arXiv:0708.2864, PoSLAT2007:337,2007

\bibitem{Chernodub_Nambu} M.N.~Chernodub, JETP Lett. {\bf 66}, 605 (1997)


\bibitem{VZ2008}
A.I.~Veselov, and M.A.~Zubkov, JHEP 0812:109 (2008) ;



\bibitem{forms}
M.I.~Polikarpov, U.J.~Wiese, and M.A.~Zubkov, Phys. Lett. B {\bf 309}, 133
(1993).


\bibitem{Montvayold}
I.~Montvay, Nucl. Phys. B {\bf 269}, 170 (1986).

\bibitem{Montvay}
 W.Langguth, I.Montvay, P.Weisz
Nucl.Phys.B277:11,1986

W. Langguth, I. Montvay (DESY) Z.Phys.C36:725,1987




\end{thebibliography}
\end{document}